\documentstyle{basi}
\begin{document}
\volume{29}
\pubyear{2001}
\title[Flat Spectrum GRB Afterglows]{Flat Spectrum Gamma Ray Burst Afterglows}
\date{Received 15 April 2001; accepted 19 June 2001}
\author[D. Bhattacharya]%
       {Dipankar Bhattacharya\thanks{e-mail:dipankar@rri.res.in} \\
        Raman Research Institute, Bangalore 560 080}
\maketitle
\begin{abstract}
Temporal behaviour of GRB afterglow light curve is derived for the case
where the electron energy distribution is relatively hard, with the 
power-law index $p$ lying between 1.0 and 2.0.  It is shown that the
expected behaviour will be the same as that for $p>2.0$ if the upper
cutoff in the electron energy distribution evolves in direct proportion 
to the bulk Lorentz factor of the blast wave.  
\end{abstract}

\section{Introduction}
Detailed observations of afterglows of Gamma Ray Bursts over the last four
years have established that they exhibit power-law broadband spectra and
power-law temporal decay of their light curve.  The generally accepted
model for the afterglow, called the fireball model, explains this emission
as being due to synchrotron emission from a relativistically expanding
blast wave which accelerates electrons to large Lorentz factors, with a
power-law energy distribution (see Piran 1999 for a review).  Recently
in several cases the light curve of the afterglow has been seen to 
undergo a break into a steeper power law, a behaviour that is expected
if the burst is beamed into a narrow solid angle (see Rhoads 1999, 2001).
Theoretical predictions for the spectral and temporal evolution have been
made in detail for both isotropic (Wijers, Rees \& M\'{e}sz\'{a}ros 1997;
Waxman 1997a,b,c; Sari, Piran \& Narayan 1998; Wijers \& Galama 1999) and
beamed (Rhoads 1997,1999; Sari, Piran \& Halpern 1999) fireballs, and these
have enjoyed considerable success in modelling the behaviour of observed 
afterglows.

Nearly all theoretical work in the literature so far assume that the 
energy distribution of the injected electrons is a power-law:
\begin{equation}
N(\gamma_{e}) \propto \gamma_{e}^{-p},\;\; 
      (\gamma_{m}<\gamma_{e}<\gamma_{u})
\end{equation}
(where $\gamma_{e}$ is the Lorentz factor of the electron and 
$\gamma_{m}$ and $\gamma_{u}$ are the lower and upper cutoff of the energy
distribution respectively), with
an index $p$ larger than $2.0$.  This assumption simplifies
the derivations, since the particles at the low-energy end dominate 
both the total number and the energy content for such a steep energy
distribution.  The values of $p$ derived from observations of most 
afterglows do indeed fall above 2.0, thereby allowing meaningful 
comparison being made between theoretical predictions and 
observations in these cases.

However, at present no compelling argument is known as to why the energy
distribution of the accelerated electrons must always be so steep.
Indeed a fairly large dispersion is seen in the spectral index distribution
of shock-accelerated electrons in Galactic shell supernova remnants,
which include several cases where $p$ is inferred to be less than 2.0
(cf.\ Green 2000).  Moreover, in Crab-like 
nebulae, where the particle energy distribution is shaped possibly
by a relativistic standing shock (Rees \& Gunn 1974, Kennel \& Coroniti
1984), the value of $p$ is almost always found to be less than 2.0.
In the context of GRB afterglows, $p \sim 1.5$ has been invoked for
GRB~000301c (Panaitescu 2001) and for GRB~010222  (Sagar et al 2001, 
Cowsik et al 2001).

Clearly, modelling of such a hard spectrum afterglow at present suffers
from the handicap that theoretical predictions specific to such energy
distributions are not available in the literature.  Panaitesu (2001)
makes a detailed case study of GRB~000301c with $p \sim 1.5$, but does
not provide general results that are easily applicable to other cases.
The aim of this paper is therefore to extend the predictions of the 
fireball model to
the case of $p < 2.0$.  In this paper I will address only the most 
commonly used spectral and dynamical regimes, namely slow cooling, adiabatic 
evolution for both isotropic and beamed bursts. I will consider a range
of $p$ between 1.0 and 2.0 and make certain simplifying assumptions that 
allow easy analytical treatment.  I will also assume that the
afterglow is optically thin over the entire range of frequencies 
of interest.
A more detailed and exhaustive study of hard spectrum fireball
models will be reported in a future publication (D. Bhattacharya \& 
K.M. Basu, in preparation).  In what follows I will adopt, wherever
applicable, expressions from Rhoads (1999) and Wijers \& Galama (1999)
with slight modification in notation.

\section{Adopted results}
I list below the expressions already available in the literature 
which I adopt for the purposes of the present paper.  These expressions
are not affected by the change of energy distribution index $p$.

\subsection{Dynamics}
I will confine myself to the adiabatic regime of expansion of the blast wave.
Rhoads (1999) presents results for dynamical evolution both before and 
after the light curve break, which corresponds to a time $t_b$ (in the
frame of the observer) when the initially tightly collimated ejecta 
begins to expand predominantly sideways.  The expressions corresponding 
to times before $t_b$ can also be used to represent isotropic bursts
by setting the initial solid angle of collimation to $4\pi$.

From Rhoads (1999) I adopt the following expressions for the dynamical 
evolution of the blast wave:
\begin{eqnarray}
\Gamma & = & 2^{-5/4} \left( \frac{3 E_0}{\pi \theta_0^2 c^2 \rho}
                      \right ) ^{1/8}
                      \left( \frac{1+z}{ct} \right)^{3/8} \;\; (t<t_b) \\
\Gamma & = & \Gamma_b \left( \frac{t}{t_b} \right)^{-1/2} \;\; (t>t_b) \\
t_{b} & = & (1+z) \left( \frac{3}{\pi} \right)^{1/3} \frac{5^{8/3}}{64}
           \frac{c}{c_s} \left( \frac{E_0}{\rho c_s^5} \right)^{1/3}
           \theta_0^2 \\
\Gamma_{b} & = & \frac{2 c_s}{5c} \frac{1}{\theta_0}
\end{eqnarray}
where $\Gamma$ is the bulk Lorentz factor of the blast wave, $\Gamma_b$
the value of $\Gamma$ at $t=t_b$, $E_0$ is the total energy of the blast 
wave, $\theta_0$ is the initial opening angle of the collimated ejecta,
$\rho$ is the ambient density and $c_s$ is the sound speed of the postshock
medium, also taken to be the speed of lateral expansion. $c$, as usual, is
the speed of light.  For a spherical blast wave, the appropriate expressions
can be obtained by substituting $\pi \theta_0^2$ with $4\pi$.  The time
$t$ is measured in the frame of the earthbound observer. $z$ is the redshift
of the afterglow.

\subsection{Magnetic Field}
Using the expressions in Rhoads (1999) the evolution of the postshock
magnetic field in the blast wave (as measured in the comoving frame)
can be written as:
\begin{equation}
B = \left( 8 \pi \frac{5c}{3c_s} \epsilon_B \rho \right)^{1/2} \Gamma c 
\end{equation}
where $\epsilon_B$ is the fraction of the postshock thermal energy 
converted into magnetic energy.

\subsection{Radiation}
As explained by Wijers \& Galama (1999), the observed location of the
peak of the synchrotron spectrum radiated by a single electron of Lorentz
factor $\gamma_{e}$ is
\begin{equation}
\nu(\gamma_{e}) = \frac{0.286}{1+z} \frac{e}{\pi m_e c} \Gamma B \gamma_e^2
\label{eq:nu}
\end{equation}
Integrated over the power-law energy distribution of electrons, one obtains
a power-law radiation spectrum for the whole afterglow, with the peak lying 
at
\begin{equation}
\nu_m = \frac{x_p}{1+z} \frac{e}{\pi m_e c} \Gamma B \gamma_m^2
\label{eq:nulowdef}
\end{equation}
where $\gamma_{m}$ is the lower cutoff of the energy distribution (see
eq.~(1)).  The factor $x_p$ is a function of the index $p$ of the energy
distribution.  For $1.0<p<2.0$ the value of $x_p$ lies between
$\sim 2.0$ and $\sim 0.65$ (Wijers \& Galama 1999). Here $e$ and $m_e$
are the charge and the mass of the electron respectively.
The received flux per unit frequency at this peak of the afterglow
spectrum is given by
\begin{equation}
F_m = \Gamma {\rm\bf N}_e \phi_p \frac{\sqrt{3} e^3 B}{m_e c^2} 
      \frac{1+z}{\Omega d^2}
\label{eq:fmdef}
\end{equation}
where ${\rm\bf N}_e$ is the total number of radiating electrons, 
$\Omega$ is the
solid angle in which the radiation is beamed and $d$ is the luminosity
distance to the afterglow from the observer. $\phi_p$ is a $p$-dependent
factor, and lies between $0.4$ and $0.6$ for $1.0<p<2.0$ (Wijers \&
Galama 1999).  According to Rhoads (1999) $F_m$ works out to be,
for $t<t_b$,
\begin{equation}
F_m = \sqrt{10\pi}\frac{\phi_p\epsilon_B^{1/2}}{\mu_e m_p}
          \frac{e^3}{m_e c^3} \sqrt{\frac{c}{c_s}}
          \frac{\rho^{1/2}E_0}{\pi\theta_0^2}
          \frac{1+z}{d^2} 
\label{eq:fm0def}
\end{equation}
where $\mu_e$ is the mean molecular weight of the ambient medium and
$m_p$ is the proton mass.  Additional factors of order unity will need
to be inserted in eqs.~(\ref{eq:nulowdef})--(\ref{eq:fm0def}) to represent
the result of integration over different parts of the fireball.
Eq.~(\ref{eq:fm0def}) shows that $F_m$ is independent of time, 
i.e. the flux at the peak is constant.  If we denote this value as 
$F_m^0$, then the evolution after $t_b$ can be written as (Rhoads 1999)
\begin{equation}
F_m = F_m^0 \left( \frac{t}{t_b} \right)^{-1}.
\label{eq:fmjetdef}
\end{equation}

The electron energy above which synchrotron cooling is important within the
expansion time corresponds to the Lorentz factor (Wijers \& Galama 1999)
\begin{equation}
\gamma_c = \frac{6\pi m_e c}{\sigma_{\rm T} \Gamma B^2 t}
\label{eq:gammacdef}
\end{equation}
where $\sigma_{\rm T}$ is the Thomson scattering cross section.

Using eq.~(\ref{eq:nu}) and the expressions of the comoving magnetic field
given above one obtains the expression for the cooling frequency from
$\gamma_c$:
\begin{eqnarray}
\nu_c & = & \frac{0.286\times 384 c^{1/2}}{(1+z)^{5/2} (40)^{3/2}}
        \frac{e m_e}{\sigma_{\rm T}^2} \left( \frac{c_s}{c} \right)^{3/2}
        \epsilon_B^{-3/2}\frac{\theta_0}{\rho E_0^{1/2}} t^{-1/2}
        \nonumber \\
 & &    (t < t_b) \\
\nu_c & = & \nu_c(t_b) = {\rm constant} \;\;\; (t>t_b) .
\end{eqnarray}

\section{Results for Flat Spectral Index}

We now have all the pieces necessary to compute the evolution of the
afterglow spectrum for a hard ($p<2$) energy distribution of electrons.
For $1<p<2$ and $\gamma_{u} \gg \gamma_{m}$ the only way this modifies
the evolution is by changing the evolution of $\gamma_{m}$ with time.
As in Sari, Piran \& Narayan (1998) we note that the postshock particle
density and energy density are $4\Gamma n$ and $4\Gamma^2 n m_p c^2$
respectively, where $n$ is the number density of the ambient medium.
Assuming a fraction $\epsilon_e$ of the postshock thermal energy goes
into power-law electrons, these quantities can be equated to integrals 
over the electron energy distribution:
\begin{eqnarray}
\int_{\gamma_{m}}^{\gamma_{u}} N(\gamma_e)d\gamma_e & = & 4\Gamma n \\
\int_{\gamma_{m}}^{\gamma_{u}} \gamma_e m_e c^2 N(\gamma_e)d\gamma_e 
     & = & \epsilon_e 4\Gamma^2 n m_p c^2
     \label{eq:energy_int}
\end{eqnarray}
Clearly, for $1<p<2$ the dominating limit in the first integral is
$\gamma_{m}$ while that in the second integral is $\gamma_{u}$.  Using the
fact that $\gamma_{m} \ll \gamma_{u}$ one then obtains
\begin{equation}
\gamma_{m} = \left[ \epsilon_e \left( \frac{2-p}{p-1} \right)
                    \frac{m_p}{m_e} \Gamma \gamma_{u}^{p-2} 
             \right]^{1/(p-1)}
\label{eq:flatgmdef}
\end{equation}
This is the key element that causes the difference of evolution between
the hard spectrum and steep spectrum afterglows. To recall (e.g. from
Sari, Piran \& Narayan 1998), for $p>2$,
the value of $\gamma_{m}$ evolves as
\begin{equation}
\gamma_{m} = \epsilon_e \left( \frac{p-2}{p-1} \right) \frac{m_p}{m_e}
             \Gamma
\label{eq:steepgmdef}
\end{equation}
The integral in eq.~(\ref{eq:energy_int}) is carried out over the 
injected energy spectrum of electrons, which is an unbroken power-law
up to $\gamma_{u}$.  This, therefore, is a measure of the total energy
the acceleration process injects into relativistic electrons, and for the
purposes of this paper I assume that this is a constant fraction 
($\epsilon_e$) of the postshock thermal energy.  The spectrum of the
{\em accumulated} electrons, however, would steepen to 
$\gamma_{e}^{-(p+1)}$ beyond the cooling break $\gamma_{c}$ 
(eq.~\ref{eq:gammacdef}) because of the radiation losses suffered after
acceleration. Over the period of interest, $\gamma_{c}$ would in general
be much less than $\gamma_{u}$ (see, e.g.\ Gallant \& Acherberg 1999,
Gallant, Achterberg \& Kirk 1999), so for $p<2$, only a small fraction
of the total injected energy will remain in the accumulated electrons.
Dai and Cheng (2001) have computed the evolution of the afterglow assuming
that the ratio ($\epsilon_{c}$) of this remaining energy to the postshock
thermal energy stays constant with time.  While the constancy of either
$\epsilon_{e}$ or $\epsilon_{c}$ as defined above is a questionable 
assumption, the degree of difficulty in arranging a physical situation
to maintain a constant $\epsilon_{c}$ is certainly greater.  We therefore
derive our results assuming a constant $\epsilon_{e}$, although the
final results will be general enough for application to either case.

Eq.~(\ref{eq:flatgmdef}) shows that the evolution of $\gamma_{m}$ in
the hard spectrum case depends on how $\gamma_{u}$ changes with time.
This depends on the details of the particle acceleration process in
the ultrarelativistic blast wave, which have so far not been very well
understood (see the review by Bhattacharjee and Sigl (2000) and references
therein).  Broadly speaking, the maximum energy achieved by an electron
in the acceleration process would be limited either by radiation losses
within the acceleration cycle time or by the cycle time itself exceeding
the age of the blast wave.  These quantities depend on the shock parameters
as well as the upstream magnetic field strength.  Since the evolution of 
most of the shock parameters can be expressed as power-law dependences
on $\Gamma$, for the purposes of this paper I make the simplifying, 
but perhaps not very unreasonable assumption that $\gamma_{u}$ for a 
given afterglow is a function of $\Gamma$ alone, and parametrize this 
dependence as a power law:
\begin{equation}
\gamma_{u} = \xi \Gamma^q
\label{eq:qdef}
\end{equation}
where $\xi$ is a constant of proportionality. The value of $q$, however,
may not be constant with time, and may depend on the dynamical regime.
For example, in the simplest acceleration models (cf. Gallant and Achterberg 
1999), $q \sim 0.5$, independent of dynamical regime, if the acceleration
is limited by radiative losses; but if the age $t$ of the blast wave limits
the acceleration, $\gamma_{u} \propto \Gamma t$, which yields a dynamics 
dependent $q$.

Eq.~(\ref{eq:qdef}) yields
\begin{equation}
\gamma_{m} = \left[ \epsilon_e \left( \frac{2-p}{p-1} \right)
                    \frac{m_p}{m_e} \xi^{p-2} 
             \right]^{1/(p-1)}
             \Gamma^{(1+pq-2q)/(p-1)}
\label{eq:flatgmq}
\end{equation}
The dependence of $\gamma_{m}$ on $\Gamma$ reduces to that for $p>2$
(eq.~\ref{eq:steepgmdef}) if $q=1$, i.e. if the upper cutoff energy
also is directly proportional to the bulk Lorentz factor of the shock.
In this case all the results derived for the temporal behaviour of 
$p>2$ afterglows will also be applicable to those with $p<2$.

It is now straightforward to obtain the dependence of $\nu_m$ on time
by inserting eq.~(\ref{eq:flatgmq}) in eq.~(\ref{eq:nulowdef}), and
using the appropriate expressions for $\Gamma$ and $B$.  The result is
\begin{eqnarray}
\nu_m & = & \frac{1}{1+z} \frac{x_p}{\pi} \frac{e}{m_e}
       \left[ \frac{40\pi}{3} \frac{c}{c_s} \epsilon_B \rho \right]^{1/2}
       \left[ \epsilon_e \left( \frac{2-p}{p-1} \right) \frac{m_p}{m_e}
              \xi^{p-2} \right]^{2/(p-1)} \times \nonumber \\
 & &   \left[ 2^{-5/4} \left( \frac{3E_0}{\pi\theta_0^2 c^2 \rho} 
              \right)^{1/8} \left( \frac{1+z}{c} \right)^{3/8}
       \right]^{2(p+pq-2q)/(p-1)} t^{-\frac{3}{4}\frac{p+pq-2q}{p-1}}
\end{eqnarray}
for $t<t_b$ and
\begin{eqnarray}
\nu_m & = & \frac{1}{1+z} \frac{x_p}{\pi} \frac{e}{m_e c}
       \left[ \epsilon_e \left( \frac{2-p}{p-1} \right) \frac{m_p}{m_e}
              \xi^{p-2} \right]^{2/(p-1)} \times \nonumber \\
 & &   B_b \Gamma_b^{(p+1+2pq-4q)/(p-1)} 
       \left( \frac{t}{t_b} \right)^{-(p+pq-2q)/(p-1)}
\end{eqnarray}
for $t>t_b$. Here $B_b$ stands for $B(t_b)$.

Noticing now that below and above the cooling break the afterglow flux
is given by
\begin{eqnarray}
F_{\nu} & = & F_m \left(\frac{\nu}{\nu_m}\right)^{-(p-1)/2}
              \;\;\; (\nu_m<\nu<\nu_c) \\
F_{\nu} & = & F_m \left(\frac{\nu_c}{\nu_m}\right)^{-(p-1)/2}
                  \left(\frac{\nu}{\nu_c}\right)^{-p/2}
              \;\;\; (\nu_c<\nu<\nu_u)
\end{eqnarray}
(where $\nu_u$ is $\nu(\gamma_u)$), we can easily obtain the
time dependence of afterglow flux by inserting the time dependence
of $F_m$, $\nu_m$ and $\nu_c$.  The final results are
\begin{eqnarray}
F_{\nu} & \propto & \nu^{-(p-1)/2} t^{-3(p+pq-2q)/8}
                    \;\;\; (\nu_m<\nu<\nu_c) \\
F_{\nu} & \propto & \nu^{-p/2} t^{-[3p+2+3q(p-2)]/8}
                    \;\;\; (\nu_c<\nu<\nu_u)
\end{eqnarray}
for $t<t_b$ and
\begin{eqnarray}
F_{\nu} & \propto & \nu^{-(p-1)/2} t^{-[p+2+q(p-2)]/2}
                    \;\;\; (\nu_m<\nu<\nu_c) \\
F_{\nu} & \propto & \nu^{-p/2} t^{-[p+2+q(p-2)]/2}
                    \;\;\; (\nu_c<\nu<\nu_u)
\end{eqnarray}
for $t>t_b$.  As one can verify, these expressions reduce to the
familiar expressions for $p>2$ by setting $q=1$.  Further, in the 
case of constant $\epsilon_{c}$ (Dai and Cheng 2001) $\gamma_{c}$ 
plays the role of $\gamma_{u}$, and the corresponding results can 
be obtained by inserting the dependence of $\gamma_{c}$ on $\Gamma$ 
in the above equations: $q=-1/3$ for $t<t_{b}$ and $q=-1$ for $t>t_{b}$.

\section{Conclusions}
I have presented above the expected behaviour of GRB afterglow light
curves when the index $p$ of the power-law energy distribution of 
electrons lies in the range $1.0<p<2.0$.  The results presented here
correspond to the optically thin, adiabatic, slow cooling regime.  The
total energy content in such an energy distribution is dominated by
the upper cutoff Lorentz factor $\gamma_{u}$, and hence the evolution
of $\gamma_{u}$ influences the evolution of the light curve.  
I derive the light curve behaviour assuming a simple power-law 
dependence of $\gamma_{u}$ on the bulk Lorentz factor $\Gamma$ of 
the blast wave.  It follows that the behaviour of the light curve 
for $1<p<2$
will be similar to that for $p>2$ if $\gamma_u \propto \Gamma$. 

It ought to be remembered that the broken power-law description of the
afterglow spectrum and light curve presented here represents only 
the asymptotic behaviour, in reality the transitions between different
regimes are expected to be smooth.  Moreover, for relatively hard
electron energy distributions considered here, synchrotron cooling is
expected to cause a pile-up of particles at the cooling break 
$\gamma_{c}$ (cf. Pacholczyk 1970) and hence some excess emission
(i.e.\ a local peak) near the cooling frequency $\nu_{c}$ may be observed.
Some of the results presented above find application in modelling
the light curve and spectrum of GRB~010222 afterglow (Sagar et al
2001, Cowsik et al 2001).

\section*{Acknowledgements}
Valuable comments by the referees have helped improve the presentation 
of the paper.  I thank K.M. Basu for discussions.


\begin{thebibliography}{99}
\bibitem{bg:00} Bhattacharjee, P., Sigl, G., 2000, Phys. Rep., 
                {\bf 327}, 109
\bibitem{cow:01} Cowsik, R. et al, 2001, BASI (in press),
                \mbox{astro-ph/0104363}
\bibitem{dc:01} Dai, Z.G., Cheng, K.S., 2001, ApJ Lett (submitted),
                \mbox{astro-ph/0105055}
\bibitem{ga:99} Gallant, Y.A., Achterberg, A., 1999,
                MNRAS, {\bf 305}, L6
\bibitem{gak:99} Gallant, Y.A., Achterberg, A., Kirk, J.G., 1999,
                 A\&AS, 138, 549
\bibitem{green:00} Green, D.A., 2000, {\em A Catalogue of Galactic 
                 Supernova Remnants (2000 August version)}, Mullard
                 Radio Astronomy Observatory, Cavendish Laboratory,
                 Cambridge, UK (available on the World Wide Web at
                 {\sf http://www.mrao.cam.ac.uk/surveys/snrs/})
\bibitem{kc:84} Kennel, C.F., Coroniti, F.V., 1984, ApJ, {\bf 283}, 710
\bibitem{pach:70} Pacholczyk, A.G., 1970, {\em Radio Astrophysics}, Freeman,
                  San Francisco
\bibitem{pana:01} Panaitescu, A., 2001, ApJ, {\bf 556}, in press,
                  \mbox{astro-ph/0102401}
\bibitem{piran:99} Piran, T., 1999, Phys. Rep., {\bf 314}, 575
\bibitem{rg:74} Rees, M.J., Gunn, J.E., 1974, MNRAS, {\bf 167}, 1
\bibitem{rhoads:97a} Rhoads, J.E., 1997, ApJ, {\bf 487}, L1
\bibitem{rhoads:99} Rhoads, J.E., 1999, ApJ, {\bf 525}, 737
\bibitem{rhoads:01} Rhoads, J.E., 2001, invited review at 9th Marcel
                    Grossman Meeting, \mbox{astro-ph/0103028}
\bibitem{sag:01a} Sagar, R. et al, 2001, BASI (in press),
                  \mbox{astro-ph/0104249}
\bibitem{spn:98} Sari, R., Piran, T., Narayan, R., 1998, ApJ, {\bf 497}, L17
\bibitem{sph:99} Sari, R., Piran, T., Halpern, J.P., 1999, ApJ, {\bf 519}, L17
\bibitem{waxman:97a} Waxman, E., 1997a, ApJ, {\bf 485}, L5
\bibitem{waxman:97b} Waxman, E., 1997b, ApJ, {\bf 489}, L33
\bibitem{waxman:97c} Waxman, E., 1997c, ApJ, {\bf 491}, L19
\bibitem{wg:99} Wijers, R.A.M.J., Galama, T.J., 1999, ApJ, {\bf 523}, 177
\bibitem{wrm:97} Wijers, R.A.M.J., Rees, M.J., M\'{e}sz\'{a}ros, P. 1997, 
                 MNRAS, {\bf 288}, L51
\end{thebibliography}
\end{document}